# Magnetic-Field-Induced Contraction in Sintered $CrSb_{0.5}Te_{0.5}$

Yoshihiko Okamoto[1,*], Yuki Kubota[2], Naoyuki Katayama[3], Kohki Takahashi[4], Daigorou Hirai[2], and Koshi Takenaka[2]

[1] *Institute for Solid State Physics, the University of Tokyo, Kashiwa 277-8581, Japan*
[2] *Department of Applied Physics, Nagoya University, Nagoya 464-8603, Japan*
[3] *Graduate School of Environmental, Life, Natural Science and Technology, Okayama University, Okayama 700-8530, Japan*
[4] *Institute for Materials Research, Tohoku University, Sendai 980-0812, Japan*
*yokamoto@issp.u-tokyo.ac.jp

The magnetic-field-induced strains and their underlying mechanisms in sintered samples of CrSb and $CrSb_{0.5}Te_{0.5}$ are investigated. Sintered CrSb exhibits negligible magnetic-field-induced strain, whereas sintered $CrSb_{0.5}Te_{0.5}$ shows a pronounced volume contraction, reaching $\Delta V/V =$ –920 ppm when a magnetic field of 9 T is applied. This magnetic-field-induced contraction occurs over a wide temperature range, extending up to room temperature, and is particularly pronounced at approximately 180 K, which is well below the Curie temperature. The large negative thermal expansion of sintered $CrSb_{0.5}Te_{0.5}$ due to the microstructural effects, reaching $\Delta V/V =$ –2900 ppm between 160 and 350 K, strongly suggests that the magnetic-field-induced contraction arises from the interplay between anisotropic lattice deformation and microstructural effects. The occurrence of magnetic-field-induced contraction instead of expansion, in contrast to other Cr-based magnets, is a distinctive feature of $CrSb_{0.5}Te_{0.5}$, in which the canted magnetic structure may play an important role.

The relationship between the magnetism and volume of solids has been extensively studied. A representative example is the spontaneous volume magnetostriction in Invar alloys and the resulting low thermal expansion. In Invar alloys, the development of ferromagnetic order upon cooling gives rise to spontaneous volume magnetostriction, and the associated volume expansion compensates for ordinary thermal contraction, resulting in exceptionally small thermal expansion. Owing to this property, Invar alloys are widely used in applications that require high dimensional stability. Metallic ferromagnets exhibiting strong spontaneous volume magnetostriction also show substantial volume changes by applying magnetic fields, known as forced volume magnetostriction [1]. For example, a volume expansion of 70 ppm under a magnetic field of 0.5 T was reported for an Invar alloy [2]. In these metallic ferromagnets, volume magneto-striction often manifests as an increase in volume upon cooling or with increasing magnetic field. This behavior can be understood in terms of the relationship between the stability of ferromagnetism and the width of the electronic band. In general, an increase in the unit-cell volume reduces the overlap between the orbitals on neighboring atoms, leading to narrower electronic bands. If this narrowing increases the density of states near the Fermi energy, itinerant ferromagnetism can become more stable. Consequently, when an applied magnetic field stabilizes the ferromagnetic state, the state with a larger volume becomes energetically favorable, resulting in magnetic-field-induced expansion.

In recent years, various magnetic-field-induced volume changes distinct from volume magnetostriction in conventional metallic ferromagnets have been observed in Cr-based magnets [3-6]. In antiferromagnets $LiInCr_4S_8$ and $AgCrS_2$, large volume expansions reaching $\Delta V/V = 700$ ppm under a magnetic field of 9 T were observed just below the antiferromagnetic ordering temperature of $T_N = 24$ K and at $T_N = 42$ K, respectively [3,4]. In these compounds, volume expansion arises from the magnetic-field-induced stabilization of an antiferromagnetically ordered phase with a larger volume than the paramagnetic phase. $ZnCr_2Se_4$ has also been reported to exhibit large magnetic-field-induced strains [7]. A common feature of all these Cr-based magnets is their strong spin–lattice coupling. The strong dependence of the magnetic interactions between Cr spins and the Cr–Cr interatomic distance is consistent with strong spin–lattice coupling [8,9].

Large magnetic-field-induced expansions have been observed over wide temperature ranges in sintered samples of ferromagnets $Cr_3Te_4$ and $Cr_2Te_3$ [5,10]. In particular, $Cr_3Te_4$ exhibits a volume expansion exceeding $\Delta V/V$ = 1000 ppm under a magnetic field of 9 T near room temperature [5]. In these sintered chromium tellurides, the interplay between the anisotropic lattice deformation associated with magnetic ordering and microstructural effects has been suggested to play an important role in producing a large magnetic-field-induced expansion [5]. Such microstructural effects arise when the anisotropic deformation of crystal grains, involving elongation along one crystallographic direction and shortening along another direction, causes changes in the sizes of voids within the sintered sample, leading to a macroscopic volume change [Fig. 1(a)]. Such microstructural effects are a well-established mechanism of negative thermal expansion in sintered samples [11,12]. However, such effects have rarely been considered as a mechanism for magnetic-field-induced volume changes.

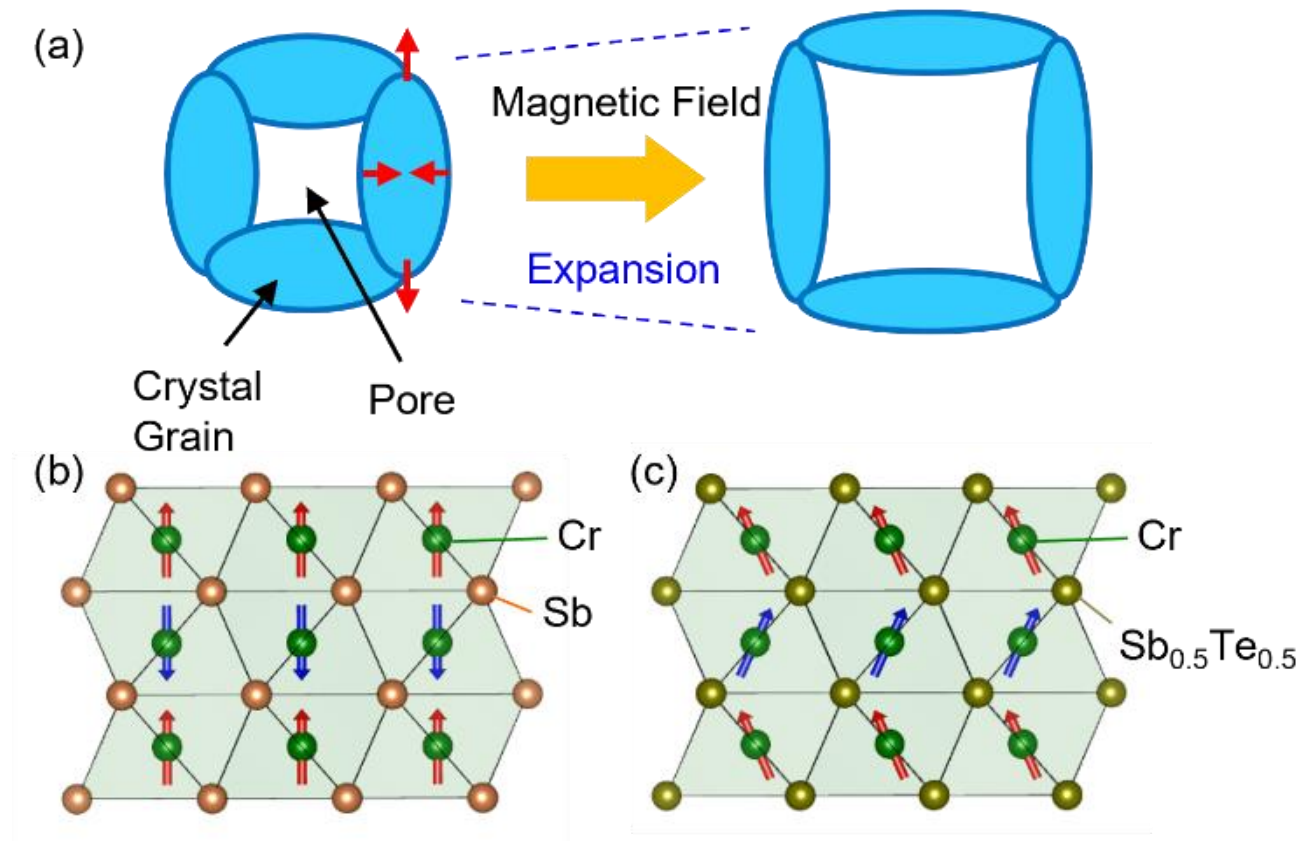

Fig. 1. (a) Schematic of magnetic-field-induced expansion of a sintered sample through microstructural effects. (b) Magnetic structure of CrSb. (c) Schematic of a possible canted magnetic structure of $CrSb_{0.5}Te_{0.5}$. The in-plane spin orientations are chosen for illustrative purposes.

In this study, we focus on CrSb and $CrSb_{0.5}Te_{0.5}$. CrSb is an antiferromagnetic metal with a hexagonal NiAs-type crystal structure and a high $T_N$ exceeding 700 K [13-15]. CrSb exhibits a collinear A-type antiferromagnetic order, in which Cr spins are aligned along the crystallographic $c$ axis and coupled antiferromagnetically between adjacent ferromagnetic layers, as shown in Fig. 1(b). In recent years, CrSb has attracted considerable attention as an altermagnet because of its large spin splitting reaching 1 eV [16-22]. Substituting Te for Sb in CrSb suppresses $T_N$ to below 300 K in $CrSb_{0.5}Te_{0.5}$ [23,24]. Importantly, $CrSb_{0.5}Te_{0.5}$ exhibits a canted ferromagnetic state with a noncollinear spin arrangement, as depicted in Fig. 1(c); however, the detailed spin arrangement remains unclear. Further Te substitution beyond 60% leads to a collinear ferromagnetic phase. In this study, dilatometric measurements reveal that sintered $CrSb_{0.5}Te_{0.5}$ exhibits a large magnetic-field-induced contraction of up to $\Delta V/V$ = –920 ppm upon application of a magnetic field of 9 T at 150–180 K. Furthermore, negative thermal expansion of up to $\Delta V/V$ = –2900 ppm is observed between 160 and 350 K. Synchrotron X-ray diffraction (XRD) measurements show that the lattice parameters of $CrSb_{0.5}Te_{0.5}$ exhibit strongly anisotropic temperature dependences, whereas the unit-cell volume increases monotonically with the temperature increasing between 110 and 400 K. XRD measurements under magnetic fields at 180 K also suggest anisotropic changes in the lattice parameters upon application of a magnetic field. These results indicate that the interplay between anisotropic lattice deformation and microstructural effects plays an important role in the emergence of both negative thermal expansion and magnetic-field-induced contraction. In particular, the large magnetic-field-induced contraction in sintered $CrSb_{0.5}Te_{0.5}$ over a wide temperature range contrasts with the magnetic-field-induced expansions previously reported for other Cr-based magnets.

Sintered samples of CrSb and $CrSb_{0.5}Te_{0.5}$ were synthesized by a solid-state reaction method [15,24]. Elemental Cr, Sb, and Te were mixed with 5% excess Cr for CrSb and in a stoichiometric ratio for $CrSb_{0.5}Te_{0.5}$. The mixtures were pressed into pellets and sealed in evacuated quartz tubes. The tubes were heated and held at 873–923 K, and then heated at 1173 K for 24 h for CrSb and at 1273 K for 48 h for $CrSb_{0.5}Te_{0.5}$. The tubes were then furnace-cooled to room temperature. The resulting samples were pulverized and sintered using a spark plasma sintering furnace (SPS Syntex) at 973 K for 1 min for CrSb and at 1073 K for 5 min for $CrSb_{0.5}Te_{0.5}$. The samples were characterized by powder XRD with Cu Kα radiation using a MiniFlex diffractometer (RIGAKU). Figure 2(a) shows the powder XRD patterns of the sintered samples of CrSb and $CrSb_{0.5}Te_{0.5}$ measured at room temperature. For each pattern, all the diffraction peaks, except for a small peak from an unidentified impurity phase in $CrSb_{0.5}Te_{0.5}$, are indexed on the basis of hexagonal unit cells with lattice parameters $a$ = 4.1288(8) Å and $c$ = 5.470(1) Å for CrSb and $a$ = 4.031(2) Å and $c$ = 5.913(3) Å for $CrSb_{0.5}Te_{0.5}$. These lattice parameters agree well with those reported in previous studies [13-15,23,24].

Magnetization was measured using a Magnetic Property Measurement System (Quantum Design). The linear strain under magnetic fields and linear thermal expansion of the sintered samples were measured using a strain gauge (KFLB, 120 Ω, Kyowa Electronic Instruments Co.) with a Cu reference [25]. Temperature control and application of magnetic fields were performed using a Physical Property Measurement System (Quantum Design). For isotropic samples, such as sintered polycrystals, the volume change $\Delta V/V$ under a magnetic field, $H$, is represented as $\Delta V/V = 2(\Delta L/L)_{\perp} + (\Delta L/L)_{//}$, where $(\Delta L/L)_{\perp}$ and $(\Delta L/L)_{//}$ are the linear strains measured perpendicular and parallel to $H$, respectively. For the linear thermal expansion of isotropic samples, the volume change is expressed as $\Delta V/V = 3(\Delta L/L)$.

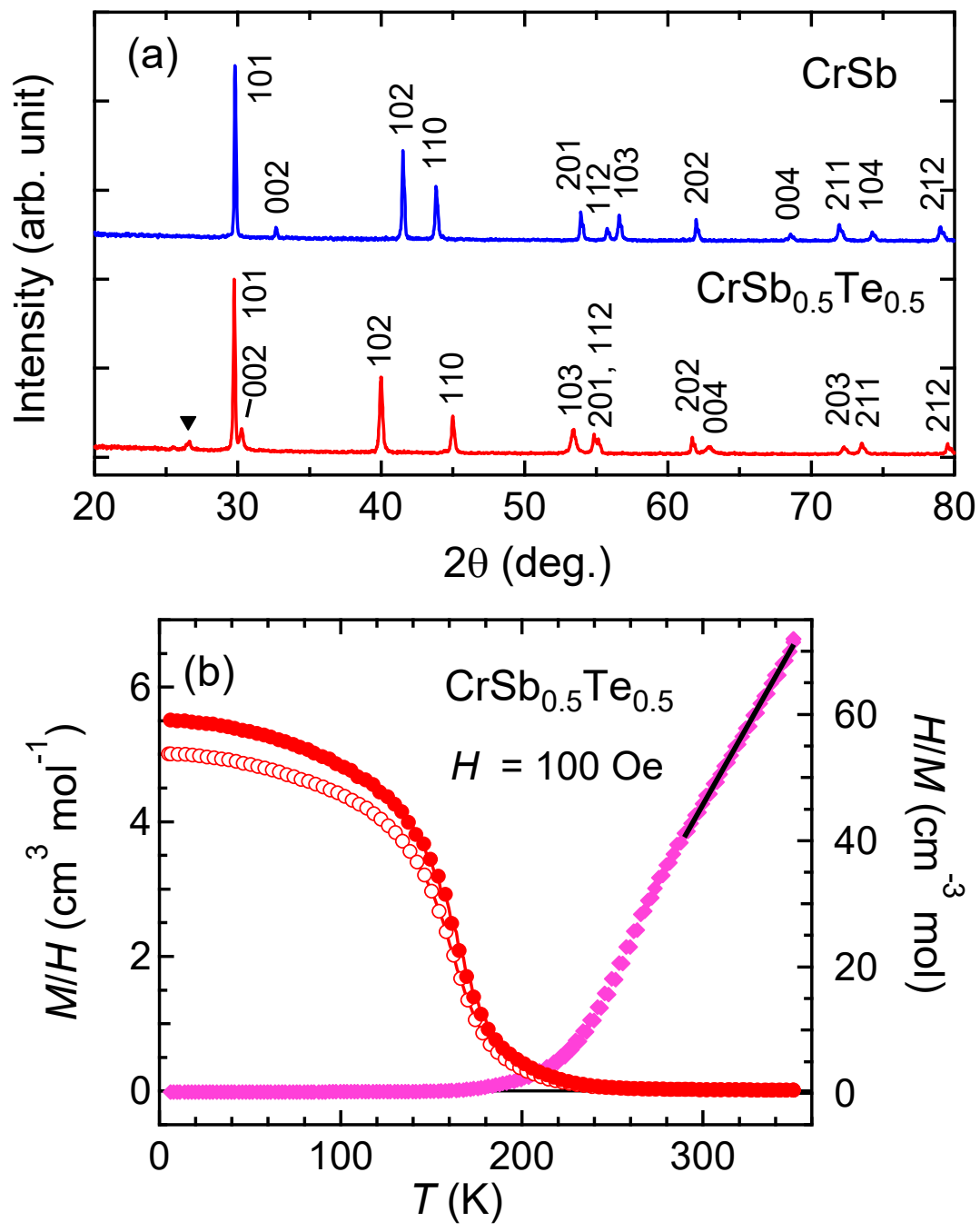

Fig. 2. (a) Schematic of magnetic-field-induced expansion of a sintered sample through microstructural effects. (b) Magnetic structure of CrSb. (c) Schematic of a possible canted magnetic structure of $CrSb_{0.5}Te_{0.5}$. The in-plane spin orientations are chosen for illustrative purposes.

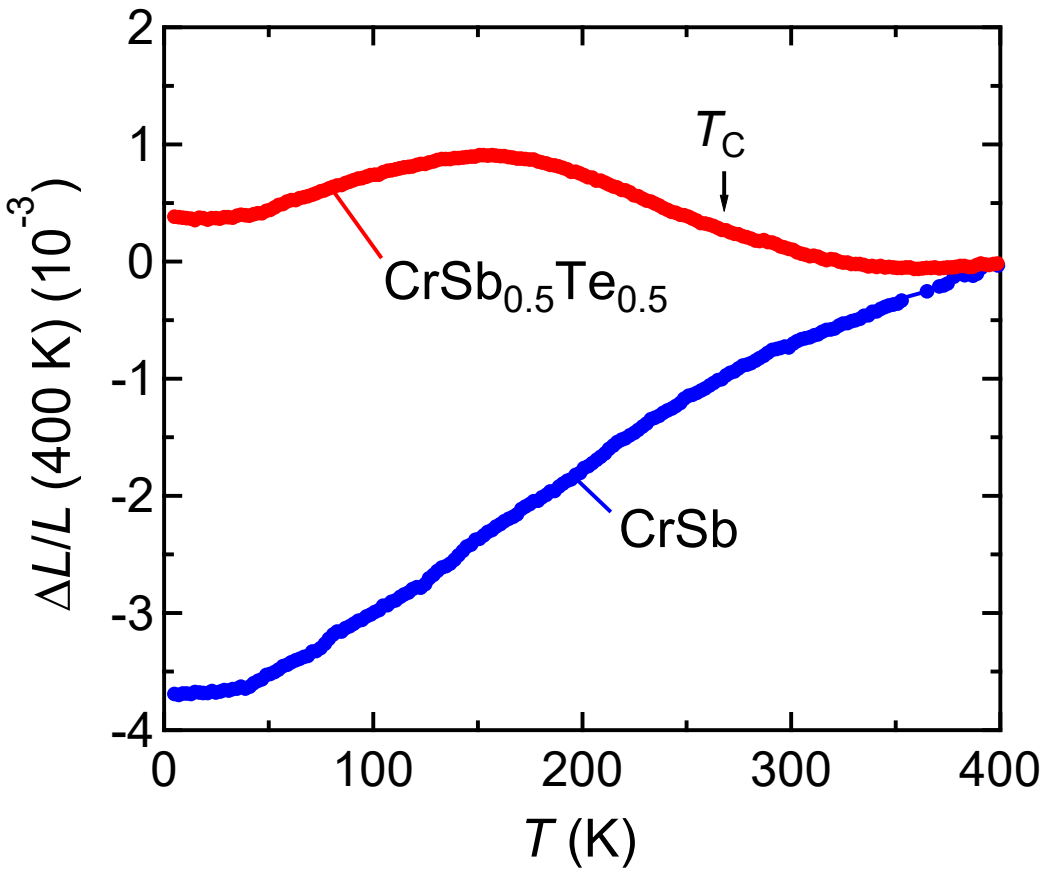

Fig. 3. Temperature dependence of linear thermal expansion of sintered samples of CrSb and $CrSb_{0.5}Te_{0.5}$ normalized to the 400 K data.

Synchrotron powder XRD measurements of sintered $CrSb_{0.5}Te_{0.5}$ at various temperatures were performed at BL5S2 of the Aichi Synchrotron Radiation Center using X-rays with a wavelength of $\lambda$ = 0.77546 Å. Powder XRD measurements of sintered $CrSb_{0.5}Te_{0.5}$ under magnetic fields of 0 and 5 T were performed using Cu K$\alpha$ radiation [26]. The crystal and spin structures were visualized using VESTA [27].

As shown in Fig. 2(b), the zero-field-cooled (ZFC) and field-cooled (FC) magnetization curves of sintered $CrSb_{0.5}Te_{0.5}$ separate below 270 K upon cooling, followed by a rapid increase in magnetization below approximately 200 K. This behavior indicates a magnetic phase transition to the canted ferromagnetic state at $T_C$ = 270 K, as reported in previous studies. The magnetization at high temperatures follows a Curie–Weiss-like behavior. Fitting the inverse magnetization data between 290 and 350 K using the Curie–Weiss law, $H/M = (T - \theta_W)/C$, yields a Curie constant of $C$ = 1.98(1) cm$^3$ K mol$^{-1}$ and Weiss temperature of $\theta_W$ = 209(3) K. The positive Weiss temperature suggests that the average magnetic interactions between Cr spins are ferromagnetic. The effective magnetic moment calculated from $C$ is $\mu_{eff}$ = 3.98 $\mu_B$/Cr, which is close to the spin-only value of $\mu_{eff}$ = 3.87 $\mu_B$/Cr for $S$ = 3/2 with a Landé g factor of $g$ = 2.

Figure 3 shows the linear thermal expansion of the sintered samples of CrSb and $CrSb_{0.5}Te_{0.5}$. CrSb exhibits positive thermal expansion over the entire temperature range below 400 K. In contrast, $CrSb_{0.5}Te_{0.5}$ exhibits large negative thermal expansion between 350 and 160 K, with a volume change reaching $\Delta V/V$ = –2900 ppm, followed by weak positive thermal expansion below 160 K. This behavior is similar to the negative thermal expansion reported for sintered $Cr_2Te_3$ and $Cr_3Te_4$, which exhibit $\Delta V/V$ = –3900 ppm between 80 and 270 K and $\Delta V/V$ = –1080 ppm between 120 and 330 K, respectively [5]. In $CrSb_{0.5}Te_{0.5}$, negative thermal expansion over a wide temperature range, including $T_C$, suggests that magnetism plays an important role in this behavior. In this temperature range, the magnetic order is not fully developed owing to thermal fluctuations and gradually develops upon cooling. As in $Cr_2Te_3$ and $Cr_3Te_4$, the development of magnetic order most likely increases the macroscopic volume of the sintered sample through microstructural effects, resulting in negative thermal expansion, as discussed below.

The sintered samples of CrSb and $CrSb_{0.5}Te_{0.5}$ exhibit markedly different magnetic-field-induced strains, as shown in Fig. 4. In Fig. 4(a), CrSb exhibits no significant strain under magnetic fields of up to 9 T. In contrast, $CrSb_{0.5}Te_{0.5}$ exhibits pronounced negative linear strains both perpendicular, $(\Delta L/L)_{\perp}$, and parallel, $(\Delta L/L)_{//}$, to the magnetic field. As a result, the sample contracts with increasing a magnetic field over the entire measured temperature range from 150 to 300 K, with a large volume decrease of $\Delta V/V$ = –920 ppm under a magnetic field of 9 T at 180 K. This $\Delta V/V$ value is among the largest reported for magnetic-field-induced volume changes in Cr-based magnets and is comparable to the forced volume magnetostriction in Invar alloys [2-5,10]. As shown in the inset of Fig. 4(c), large contractions occur at temperatures below $T_C$. The most distinctive feature of the magnetic-field-induced contraction in sintered $CrSb_{0.5}Te_{0.5}$ is that such a large volume change occurs as a contraction rather than an expansion. Magnetic-field-induced contractions of this magnitude are rarely observed in crystalline solids.

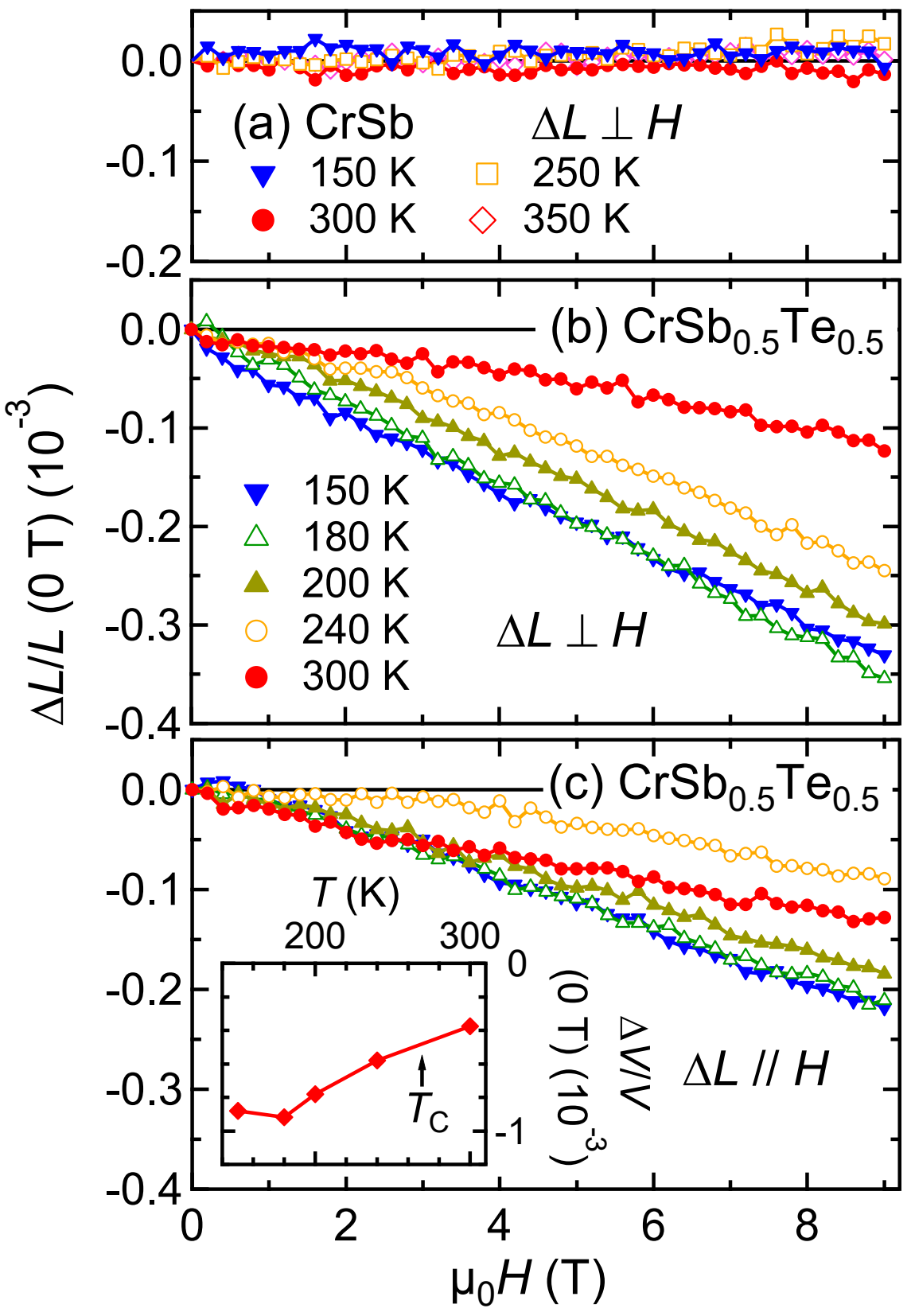

Fig. 4. Linear strains of sintered samples of (a) CrSb and (b, c) $CrSb_{0.5}Te_{0.5}$ measured in magnetic fields. All data are normalized to the zero-field data at each temperature. Panels (a, b) and (c) show the transverse and longitudinal magnetostrictions, $(\Delta L/L)_{\perp}$ and $(\Delta L/L)_{//}$, where linear strains are measured perpendicular and parallel to the magnetic field, respectively. The inset shows the volume changes in sintered $CrSb_{0.5}Te_{0.5}$ by applying a magnetic field of 9 T, estimated from the linear strain data in the main panels.

We now discuss the mechanism responsible for the large magnetic-field-induced contraction in sintered $CrSb_{0.5}Te_{0.5}$. The discrepancy between the temperature dependence of the crystallographic unit-cell volume determined by XRD and the macroscopic volume change of the sintered sample determined from dilatometric measurements is notable. As shown in Fig. 5(a), the *a*-axis length decreases monotonically when the temperature increases between 100 and 400 K, whereas the *c*-axis length increases monotonically. The changes between 100 and 400 K are $\Delta a/a = -0.4\%$ and $\Delta c/c = 2.3\%$, respectively. Because the increase in the *c*-axis length outweighs the decrease in the *a*-axis length, the unit-cell volume *v* increases with increasing temperature, exhibiting positive thermal expansion with $\Delta v/v = 1.5\%$ between 100 and 400 K. This result indicates that the negative thermal expansion observed in the dilatometric measurements originates not from a decrease in the unit-cell volume itself but from microstructural effects in the sintered sample. Such microstructural effects are well established as a mechanism of negative thermal expansion in sintered materials, including β-eucryptite, $Ca_2RuO_4$, β-$Cu_{1.8}Zn_{0.2}V_2O_7$, and $Zn_{2-x}Cu_xP_2O_7$ [28-31].

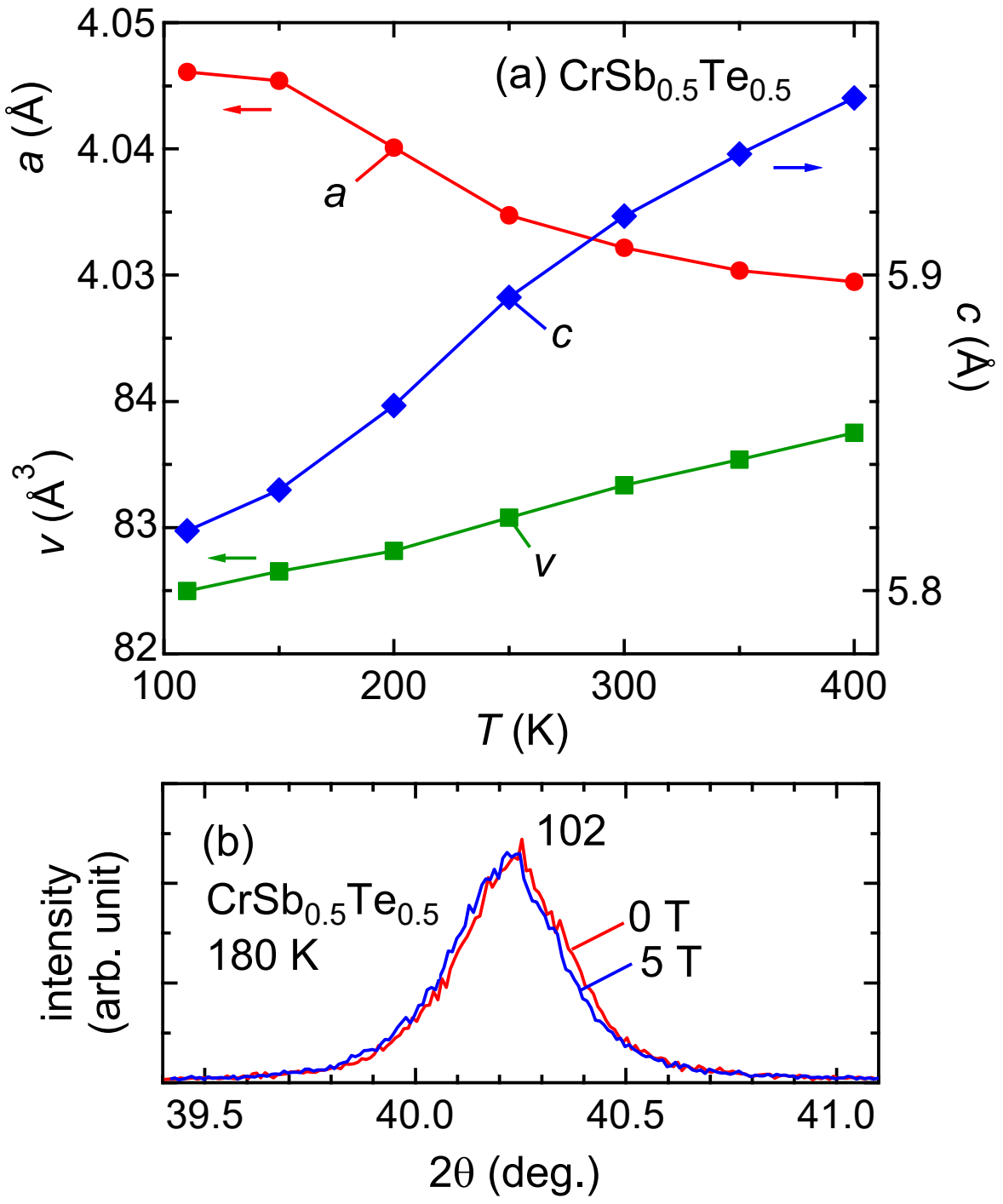

Fig. 5. (a) Temperature dependences of lattice parameters of sintered $CrSb_{0.5}Te_{0.5}$ estimated using powder synchrotron XRD data. (b) Powder XRD patterns around the 102 diffraction peak of sintered $CrSb_{0.5}Te_{0.5}$ measured under magnetic fields of 0 and 5 T at 180 K. The diffraction angles (2θ) were corrected based on the results of XRD measurements of Si under magnetic fields.

$Cr_2Te_3$ and $Ti_2O_3$ are particularly similar to $CrSb_{0.5}Te_{0.5}$. All three compounds have hexagonal crystal structures related to the NiAs type. In these compounds, the *a*-axis length decreases with increasing temperature, whereas the *c*-axis length increases by a larger proportion, resulting in an increase in the unit-cell volume *v* [5,32,33]. Nevertheless, the macroscopic volume of their sintered samples exhibits negative thermal expansion. This negative thermal expansion most likely arises from the anisotropic deformation of the crystal grains, which contract in the *ab* plane and expand along the *c* axis with increasing temperature, accompanied by changes in the shape and size of the voids within the sintered samples. For $Cr_2Te_3$ and $Cr_3Te_4$, the relationship between such microstructural effects and large magnetic-field-induced expansions has been discussed [5]. These similarities suggest that microstructural effects also play an important role in the large magnetic-field-induced contractions observed in sintered $CrSb_{0.5}Te_{0.5}$.

An intriguing feature of sintered $CrSb_{0.5}Te_{0.5}$ is the coexistence of negative thermal expansion arising from microstructural effects and large magnetic-field-induced contractions. In sintered $Cr_2Te_3$ and $Cr_3Te_4$, applying a magnetic field near $T_C$ and cooling from temperatures near $T_C$ suppress spin fluctuations, resulting in an increase in the sample volume. These changes manifest as magnetic-field-induced expansion and negative thermal expansion, respectively [5]. Although microstructural effects are involved in these sintered samples, their macroscopic volume responses resemble the forced and spontaneous volume magnetostriction in conventional metallic ferromagnets: the sample volume increases upon the application of a magnetic field and upon cooling.

Sintered $CrSb_{0.5}Te_{0.5}$ exhibits negative thermal expansion upon cooling from temperatures near $T_C$, as in the cases of $Cr_2Te_3$ and $Cr_3Te_4$. However, the application of a magnetic field induces a large decrease in the sample volume. To investigate the origin of this unusual behavior, we performed powder XRD measurements at 180 K under magnetic fields of 0 and 5 T. Small shifts in the diffraction peak positions are observed upon application of the magnetic field, as shown in Fig. 5(b). Since no peak splitting or broadening is observed, we estimate the changes in the lattice parameters by assuming that the hexagonal symmetry is preserved under the magnetic field. From these peak shifts, the *a*-axis length is estimated to decrease by approximately 0.001 Å ($\Delta a/a = -0.02\%$) and the *c*-axis length to increase by approximately 0.006 Å ($\Delta c/c = 0.1\%$) under a magnetic field of 5 T. Notably, these changes in the lattice parameters are in the same directions as those observed upon warming from 180 K, rather than those observed upon cooling in $Cr_3Te_4$ and $Cr_2Te_3$. Thus, the anisotropic lattice deformation indicated in the XRD measurements under a magnetic field is consistent with the interpretation that the magnetic-field-induced contraction arises from the same microstructural effects as those responsible for the negative thermal expansion.

These results demonstrate that, unlike in $Cr_2Te_3$ and $Cr_3Te_4$, cooling and the application of a magnetic field induce different lattice deformation in $CrSb_{0.5}Te_{0.5}$. This difference may be related to the noncollinear magnetic structure with canted spins [Fig. 1(c)] [23,24]. In such a magnetic structure, applying a magnetic field near $T_C$ is expected to not only suppress spin fluctuations but also induce a continuous evolution of the spin configuration toward a more collinear arrangement. Such magnetic-field-induced evolution of the spin configuration may produce anisotropic lattice deformations, characterized by a slight decrease in the *a*-axis length and an increase in the *c*-axis length, resulting in a large magnetic-field-induced contraction of the sintered sample through microstructural effects.

In summary, we have demonstrated that sintered $CrSb_{0.5}Te_{0.5}$ exhibits a large magnetic-field-induced contraction of up to $\Delta V/V = -920$ ppm upon application of a magnetic field of 9 T. This contraction is observed over the entire measured temperature range from 150 to 300 K, whereas sintered CrSb exhibits no significant magnetic-field-induced strain. Notably, the large magnetic-field-induced volume change in $CrSb_{0.5}Te_{0.5}$ occurs as a contraction rather than an expansion, as previously reported for other Cr-based magnets. Sintered $CrSb_{0.5}Te_{0.5}$ also exhibits a large negative thermal expansion associated with the development of magnetic order and microstructural effects, as supported by temperature-dependent XRD measurements. XRD measurements under a magnetic field indicate that the anisotropic lattice deformation induced by a magnetic field occurs in the same direction as that observed upon warming, indicating that microstructural effects are responsible for both the magnetic-field-induced contraction and the negative thermal expansion. The canted magnetic structure of $CrSb_{0.5}Te_{0.5}$ may play an important role in the occurrence of magnetic-field-induced contraction rather than expansion, in contrast to other Cr-based magnets. This study further highlights a wide variety of magnetovolume phenomena arising from the coupling between magnetism and the crystal lattice in Cr-based magnets.

## ACKNOWLEDGMENTS

We are grateful to T. Kanematsu and T. Yajima for helpful discussions. This work was partly carried out at HFLSM under the GIMRT Program of the Institute for Materials Research, Tohoku University (Proposal No. 202602-HMKGE-0063). and was partly supported by JSPS KAKENHI (Grant Nos. 23H01831, 23K26524, and 26H02231), and JST ASPIRE (Grant No. JPMJAP2314). The synchrotron X-ray diffraction experiments were performed at BL5S2 of the Aichi Synchrotron Radiation Center, Aichi Science and Technology Foundation, Aichi, Japan (Proposal No. 202205005).